\documentclass[aps,prb,twocolumn,superscriptaddress,notitlepage]{revtex4-2}
\usepackage{amsmath}
\usepackage{mathptmx}
\usepackage{newtxmath}
\usepackage[latin9]{inputenc}
\usepackage{color}
\usepackage{bm}
\usepackage{graphicx}
\usepackage[unicode=true,
 bookmarks=false,
 breaklinks=false,pdfborder={0 0 1},backref=false,colorlinks=true]
 {hyperref}
\hypersetup{
 linkcolor=blue,urlcolor=blue,citecolor=blue}

\makeatletter

\usepackage{newtxtext}
\usepackage{xspace}
\usepackage{amsbsy}
\usepackage{bm}
\usepackage{bbold}
\usepackage{graphicx}
\usepackage{color}
\usepackage{xcolor}
\usepackage{epsfig}
\usepackage{rotate}
\usepackage{fancyhdr}
\usepackage{soul}
\usepackage{physics}
\usepackage{mathrsfs}

\allowdisplaybreaks[2]

\makeatother

\begin{document}
\title{Classical spin dynamics based on SU($N$) coherent states}

\author{Hao~Zhang}
\affiliation{Department of Physics and Astronomy, The University of Tennessee,
Knoxville, Tennessee 37996, USA}
\affiliation{Materials Science and Technology Division, Oak Ridge National Laboratory, Oak Ridge, Tennessee 37831, USA}

\author{Cristian~D.~Batista}
\affiliation{Department of Physics and Astronomy, The University of Tennessee,
Knoxville, Tennessee 37996, USA}
\affiliation{Neutron Scattering Division and Shull-Wollan Center, Oak Ridge
National Laboratory, Oak Ridge, Tennessee 37831, USA}

\begin{abstract}
We introduce a classical limit of the dynamics of quantum spin systems
based on coherent states of  SU($N$), where $N$ is the dimension of the local Hilbert space. This approach, which  generalizes the well-known Landau-Lifshitz dynamics from SU(2) to SU($N$), provides a better approximation to the exact quantum dynamics for a large class of realistic spin Hamiltonians, including $S \geq 1$
systems with large single-ion anisotropy and  weakly-coupled multi-spin units, such as dimers or trimers. We illustrate this idea by comparing the spin structure factors of a  single-ion $S=1$ model that are obtained with the SU(2) and SU(3) classical spin dynamics  against the exact solution.
\end{abstract}
\maketitle

\section{Introduction}

The concept of a ``coherent state'' was first proposed by Schr\"odinger in 1926~\citep{1926Schrodinger}. He derived the state as the most classical (i.e., minimum uncertainty $\Delta x\Delta p =\hbar/2$) state of a quantum harmonic oscillator. The first practical application of this concept was introduced by Glauber and Sudarshan in 1963~\citep{glauber1963a,glauber1963b,Glauber1963c,sudarshan1963equivalence} to describe the maximal coherent light state in quantum optics, hence the name \emph{coherent state}. Coherent states were originally constructed as  eigenstates of the annihilation operator and they were generated  by applying a displacement operator of the Heisenberg-Weyl group to the vacuum state of the harmonic oscillator. In his third seminal paper~\citep{Glauber1963c}, Glauber pointed out that there are three equivalent definitions of coherent states: (i) the eigenstate of the annihilation operator, (ii) the state obtained by applying a displacement operator of the Heisenberg-Weyl group on the vacuum state of the harmonic oscillator, and (iii) the state with minimum Heisenberg uncertainty. Ten years later, Perelomov~\citep{perelomov1972} and Gilmore~\citep{Gilmore1972391} observed that the second definition of Glauber can be extended to an arbitrary Lie group. This generalization was later considered by Yaffe to explore the classical limit of quantum systems~\citep{Yaffe}. 
Moreover, using this generalization of coherent states, one can extend and prove  Dirac's conjecture---the commutator between two quantum operator times $1/(i\hbar)$ is replaced by the Poisson bracket in the classical limit---for  general Lie groups, where the Poisson bracket is defined on the classical phase space defined by the orbit of  coherent states.

Classical limits of $N$-level quantum systems based on SU($N$) coherent states have been exploited in different areas of physics~\cite{klauder85,Feng94}, ranging from atomic physics~\cite{Gilmore1972,Glauber76,Gnutzmann_1998,Drummond07,Barry08} to quantum chromodynamics~\cite{Brown86,Dickens88}.
The main goal of this work is to present a comprehensive discussion of the classical dynamics of a \emph{quantum spin system} based on SU($N$) coherent states. To understand our motivation, we must first recognize that  there are two widely used tools for modeling spin dynamics: the spin-wave theory (SWT) and the Landau-Lifshitz dynamics (LLD). The traditional SWT is a semi-classical approach, whose starting point is a mean-field state consisting of a direct product of SU(2) coherent states~\cite{LLD_DPLandau_review,LLD_Chalker_frustrate,LLD_AS_honeycomb,LLD_AS_kitaev,LLD_AS_ML,LLD_SLin_skyrmion,LLD_SZhang}. 
Quantum effects are incorporated order by order via a $1/S$ expansion~\cite{mattis2012theory,auerbach2012interacting,largeS_RMP,Anderson_largeS,Kubo_largeS,HP_expansion}. The LLD was first introduced by Landau and Lifshitz~\citep{LandauLifshitz} to describe the precession of the  magnetization in a solid. In this classical approximation, the state of the system is approximated by a  direct product of SU(2) coherent states at any time $t$. In the absence of damping, the classical LLD  equations can be obtained by taking the classical limit of the Heisenberg equation following Dirac's prescription. 

While SWT and LLD become exact in the large-$S$ limit, their  applicability to real spin systems (with a finite value of $S$) is not just limited by the presence of large quantum fluctuations. As we will discuss in this work, there are large classes of realistic spin models whose spin dynamics is not well described by the traditional SWT or LLD, but still admit an accurate  semi-classical or classical treatment. This apparent paradox disappears when we recognize that there is more than one way of taking the classical limit of a spin system~\cite{Gnutzmann_1998}.

The existence of multiple classical limits is tied to different choices of the Lie group that defines the orbit of coherent 
states~\cite{perelomov1972,Yaffe,Gitman_1993,Gnutzmann_1998}. For instance, for a three-level system ($S=1$) we can either use a mean field state that is a direct product of SU(2) coherent states or one that is a direct product of SU(3) coherent states. The second of these is the starting point of the so-called generalized SWT~\citep{muniz2014generalized}, which has been successfully applied to several models and quantum magnets~\citep{Batista04,Zapf06,Kohama11,Wierschem12,Zhang13,Zapf14,Wang17,bai2021hybridized,do2020decay}. An obvious advantage of the use of SU(3) coherent states for $S=1$ systems is that they allow to describe local quadrupolar and dipolar moments (and their fluctuations) on an equal foot.~\citep{muniz2014generalized} In contrast, SU(2) coherent states can only account for dipolar ordering and dipolar fluctuations.  



The SU($N$) generalization of the semiclassical $1/S$ expansion~\cite{do2020decay}, suggests  that the classical limit  or LLD dynamics  must also be generalized. However, to the best of our knowledge, the application of the classical SU($N$) dynamics to spin systems has not been discussed in the literature. Although the classical limit of a quantum theory based on  SU($N$) coherent states (for $N>2$) was discussed in Refs.~\citep{Gnutzmann_1998,Gitman_1993}, these works were mainly focused on mathematical aspects  of the SU($N$) coherent states. After recognizing that quantum theories of $N$-level systems admit more than one classical limit, it becomes relevant to ask what is the most adequate choice for each particular application.
In this work we focus on applications to spin Hamiltonians. Our main motivation is to demonstrate that the above-mentioned generalization of Dirac's conjecture leads to a natural generalization of the LLD, that can be used to model the excitation spectrum of large classes of quantum magnets for which the traditional LLD dynamics is simply inadequate.

This paper is organized as follows. In Sec.~\ref{sec:point1D} we review the properties of the coherent states of the Heisenberg-Weyl group, which consists of translation operators of a quantum mechanical point particle in phase space. In particular, we include the demonstration of Dirac's conjecture that was sketched by Yaffe in Ref.~\citep{Yaffe}.  The inclusion of the intermediate steps of this demonstration is useful to motivate  the general discussion on the classical limit of $N$-level quantum spin systems. In Sec.~\ref{sec:SUN_group} we review the representation theory of the SU($N$) group and the construction of the SU($N$) coherent states for degenerate representations. This section includes the mathematical background that is needed for the later sections. The classical limit of a quantum spin system based on  SU($N$) coherent states is discussed in Sec.~\ref{sec:SUN_classical}, where we show that the traditional classical limit $\hbar\rightarrow0$ is equivalent to the limit in which the dimension of the local Hilbert space of the system going to infinity. In Sec.~\ref{sec:classical_eom} we use the generalized Dirac's conjecture to write down explicitly the classical equations of motion of the SU(2) and SU(3) generators. The generalization for SU($N$) generators is discussed in the second part of Sec.~\ref{sec:classical_eom}. In Sec.~\ref{sec:sion}, we use a simple single-ion model to illustrate the importance of generalizing the LLD. The final conclusions are discussed in  the Sec~\ref{conc}. 

\section{Coherent states for a point particle in one dimension\label{sec:point1D}}

Together with the identity $\hat{1}$, the position $\hat{x}$ and momentum $\hat{p}$ operators  of a point particle in one dimension generate the so-called Heisenberg-Weyl group $H_{3}$. Correspondingly, 
these three operators form a basis of the Lie algebra $h_{3}$, whose structure is determined by the canonical
commutation relations $[\hat{x},\hat{p}]=i\hbar$, and $[\hat{x},\hat{1}]=[\hat{p},\hat{1}]=0$.
The identity, the creation and annihilation operators,
\begin{equation}
\hat{a}^{\dagger} = \sqrt{\frac{1}{\hbar}}\left( \hat{x}-i\hat{p} \right),\ \
\hat{a}           = \sqrt{\frac{1}{\hbar}}\left( \hat{x}+i\hat{p} \right),
\end{equation}
provide an alternative basis of $h_{3}$. The \emph{base state} $|0\rangle$ for coherent states is defined by the condition
\begin{equation}
\hat{a}|0\rangle=0.
\end{equation}
A coherent state is obtained by applying an element of $H_{3}$ to the base state~\citep{Glauber1963c}
\begin{equation}
| \alpha \rangle = e^{\alpha\hat{a}^{\dagger}-\bar{\alpha}\hat{a}} |0\rangle,
\label{eq:displacement}
\end{equation}
where $e^{\alpha\hat{a}^{\dagger}-\bar{\alpha}\hat{a}}\in H_{3}$ is the so-called  displacement operator~\citep{ZhangRMP} and $\alpha,\bar{\alpha}\in\mathbb{C}$.
By convention, one can use two real numbers $p$ and $q$ with $\alpha=q+ip$
to label a coherent state. 
The wave function of $|\alpha\rangle=|p,q\rangle$ corresponds to the ground state  of a simple harmonic oscillator centered around $q$
\begin{equation}
|\langle x|p,q \rangle|^{2} = (\pi\hbar)^{-1/2} \exp\left\{ (1/\hbar)\left[-(x-q)^{2}\right] \right\}.
\end{equation}
The manifold of coherent states forms an over-complete basis of the Hilbert space of a point particle in one-dimension~\citep{klauder1985coherent}.
The over-completeness is manifested by the finite overlap between two coherent states 
\begin{align}
|\langle p,q|p',q'\rangle|^{2} & = \exp\left\{ -(1/2\hbar)\left[(p-p')^{2}+(q-q')^{2} \right]\right\} 
                                   \nonumber \\
                               & =\exp \left\{ -(1/2\hbar)(\alpha-\alpha')(\bar{\alpha}-\bar{\alpha}')
                                       \right\}.
\label{eq:overlap}
\end{align}

The symbol without the hat denotes the expectation value of the corresponding operator for a coherent state:
\begin{equation}
A(p,q) \equiv \langle p,q | \hat{A} |p,q \rangle.
\label{eq:expA}
\end{equation}
Similarly, the expectation value of the product of two operators $\langle p,q | \hat{A}\hat{B} |p,q \rangle$ is given by
\begin{align}
(AB)(p,q) & = \int \frac{dp'dq'}{2\pi\hbar}
             | \langle p,q|p',q' \rangle |^{2}\nonumber \\
          & \times \frac{\langle p,q|\hat{A}|p',q'\rangle}
                  {\langle p,q|p',q'\rangle}
                  \frac{\langle p',q'|\hat{B}|p,q\rangle}
                  {\langle p',q'|p,q\rangle},
\label{eq:expvalAB}
\end{align}
where we inserted the resolution of identity 
\begin{equation}
\hat{I} = \int\frac{dp'dq'}{2\pi\hbar} | p,q \rangle \langle p,q |.
\end{equation}
Dirac's conjecture can be proved by taking the $\hbar\rightarrow0$ limit of Eq.~\eqref{eq:expvalAB}~\citep{Yaffe}. 
According to Eq.~\eqref{eq:overlap}, the first factor of the integrand of Eq.~\eqref{eq:expvalAB}
is a Gaussian that has a sharp peak at $p'=p$ and $q'=q$ as $\hbar\rightarrow0$.
Moreover, notice that for fixed $p$ and $q$, the second factor is an analytical function of $\alpha'=(q'+ip')$ and the third factor is an analytical function of $\bar{\alpha}'=(q'-ip')$. 
Therefore, we can expand the second and the third factors up to  quadratic order in $(\alpha'-\alpha)$
and $(\bar{\alpha}'-\bar{\alpha})$, respectively, 
\begin{align}
\frac{\langle p,q | \hat{A} |p',q' \rangle}{\langle p,q|p',q' \rangle} 
 & = A(\alpha) + \frac{dA}{d\alpha'}\bigg|_{\alpha'=\alpha} (\alpha'-\alpha)\nonumber \\
 & + \frac{1}{2} \frac{d^{2}A}{d\alpha'^{2}}\bigg|_{\alpha'=\alpha}
   (\alpha'-\alpha)^{2} + \mathcal{O}\big[(\alpha'-\alpha)^{3}\big],
\label{eq:papp}
\end{align}
\begin{align}
\frac{\langle p',q'| \hat{B} |p,q \rangle}{\langle p',q'|p,q \rangle}
 & = B(\alpha)+\frac{dB}{d\bar{\alpha}'}\bigg|_{\bar{\alpha}'
   = \bar{\alpha}}(\bar{\alpha}'-\bar{\alpha})\nonumber \\
 & + \frac{1}{2}\frac{d^{2}B}{d\bar{\alpha}'^{2}}\bigg|_{\bar{\alpha}'
   =\bar{\alpha}}(\bar{\alpha}'-\bar{\alpha})^{2} +\mathcal{O}\big[(\bar{\alpha}'-\bar{\alpha})^{3}\big].
\label{eq:ppbp}
\end{align}
By combining the results, we have
\begin{align}
(AB)(p,q) & \simeq\int\frac{d\alpha d\bar{\alpha}}{2\pi\hbar}|\langle\alpha|\alpha'\rangle|^{2}\bigg\{ A(\alpha)B(\alpha)\nonumber \\
 & +\frac{dA}{d\alpha}\frac{dB}{d\bar{\alpha}}(\alpha'-\alpha)(\bar{\alpha}'-\bar{\alpha})+\mathcal{L} \bigg \},
 \label{eq:abexpand}
\end{align}
where $\mathcal{L}$ includes terms (up to quadratic order) that vanish after the integration.
After computing the Gaussian integrals and keeping contributions up to first order in $\hbar$, we obtain 
\begin{equation}
(AB)(p,q) \simeq A(p,q)B(p,q)+\frac{\hbar}{2}\frac{dA}{d\alpha}\frac{dB}{d\bar{\alpha}}.
\end{equation}
In the $\hbar\rightarrow0$ limit, we have
\begin{equation}
\lim_{\hbar\rightarrow0}(AB)(p,q) =a(p,q)b(p,q),\label{eq:factorial}
\end{equation}
where
\begin{equation}
a(p,q)\equiv\lim_{\hbar\rightarrow0}A(p,q),\quad b(p,q)\equiv\lim_{\hbar\rightarrow0}B(p,q).\label{eq:expcl}
\end{equation}
Note that the functions $a(p,q)$ and $b(q,p)$ are assumed to remain finite in the $\hbar\rightarrow0$ limit. They are the so-called \emph{classical operators}~\citep{Yaffe}, that resemble the operators $\hat{A}$ and $\hat{B}$ in the classical limit.
The relation \eqref{eq:factorial} gives the \emph{factorization rule} for the expectation value of the product of two operators in the $\hbar\rightarrow0$ limit.

Finally, let us replace $\hat{B}$ with the Hamiltonian operator $\hat{H}$ and consider the $\hbar\rightarrow0$
limit of the expectation value of the right hand side of the Heisenberg equation (HE) of motion~\footnote{Note that for the complex variable $\alpha=p+iq$, ${\rm d}/{\rm d}\alpha=(\partial/\partial p-i\partial/\partial q)/2$.}
\begin{align}
\lim_{\hbar\rightarrow0}-\frac{i}{\hbar}\big[A,H\big](p,q) & =-\lim_{\hbar\rightarrow0}\bigg[\frac{\partial A}{\partial p}\frac{\partial H}{\partial q}-\frac{\partial A}{\partial q}\frac{\partial H}{\partial p}\bigg]\nonumber \\
 &=\frac{\partial a}{\partial q}\frac{\partial h}{\partial p}-\frac{\partial a}{\partial p}\frac{\partial h}{\partial q}\nonumber\\
 &=\big\{ a(p,q),h(p,q)\big\}_{PB},
\end{align}
where $h(p,q)=\lim_{\hbar\rightarrow0}\langle p,q|\hat{H}|p,q\rangle$ is the classical Hamiltonian. 
After taking the same expectation value of the left hand side of the HE, we obtain
\begin{equation}
\lim_{\hbar\rightarrow0}\langle p,q|\frac{d\hat{A}}{dt}|p,q\rangle=\frac{da(p,q)}{dt}=\big\{ a(p,q),h(p,q)\big\}_{PB}.
\end{equation}
This completes the proof of Dirac's conjecture.
Here the coherent states of $H_{3}$ play an important role in linking the quantum and classical theories for a point particle. 
As we already mentioned, the  coherent states of $H_3$ can be generalized to any Lie group~\citep{perelomov1972,Gilmore1972391}, implying that we can take the classical limit of a given quantum theory by introducing a manifold of coherent states of an appropriate Lie group~\citep{Yaffe}. 
In the rest of this paper, we focus our discussion on the SU($N$) group.

\section{Review of the SU($N$) group\label{sec:SUN_group}}
\subsection{Representation theory of SU($N$)}

In this section we review the representation theory of the Lie group SU($N$) defined as the set of $N\times N$ unitary matrices with determinant one and with the matrix multiplication as the group operation~\citep{hall2015lie}. 
This group arises naturally in the description of an $N$-level quantum-mechanical system as the set of all unitary basis transformations with determinant equal to one. 
The  Lie algebra $\mathfrak{su}(N)$, that is defined as the tangent space at the identity element of SU($N$), is a vector space over $\mathbb{C}^{N}$ of dimension  $N^{2}-1$.
In the fundamental representation, the generators (bases) of $\mathfrak{su}(N)$ are represented by the following matrices
\begin{equation}
\hat{g}_{ij}\ (i\neq j),\ \hat{H}_{1}=\frac{1}{2}(\hat{g}_{11}-\hat{g}_{22}),\ldots\ \hat{H}_{N-1}=\frac{1}{2}(\hat{g}_{N-1N-1}-\hat{g}_{NN}),
\label{eq:sun_basis}
\end{equation}
where
\begin{equation}
\hat{g}_{ij}\equiv|i\rangle\langle j|,\ i,j=1,2,\ldots N
\end{equation}
and
\begin{equation}
|i\rangle=(0,\ \ldots,\ \overset{i\text{th}}{1},\ldots,0)^{T}
\end{equation}
represents the standard basis of $\mathbb{C}^{N}$. The matrices satisfy the following commutation relations
\begin{equation}
\big[\hat{g}_{ij},\hat{g}_{kl}\big]=\delta_{kj}\hat{g}_{il}-\delta_{il}\hat{g}_{kj},\label{eq:commu1-1}
\end{equation}
and
\begin{equation}
\big[\hat{H}_{k},\hat{H}_{l}\big]=0,\ k,l=1,\ldots,N-1,\label{eq:commu2-1}
\end{equation}
where the set of $N-1$ generators $\{\hat{H}_{k}\}$ with $k=1,\ldots,N-1$ spans the Cartan subalgebra (maximal commutative subalgebra~\citep{hall2015lie}), and the remaining $N(N-1)$ generators $\hat{g}_{ij}$ are the so-called raising (lowering) operators if $i<j$ ($i>j)$. 
It is important to keep in mind that any basis of matrices that obey Eqs.~\eqref{eq:commu1-1} and ~\eqref{eq:commu2-1}
provides a set of generators of $\mathfrak{su}(N)$. 

Consider a general irreducible representation (irrep) of $\mathfrak{su}(N)$ on the vector space $V$. 
The \emph{highest-weight state} $|\mu\rangle\in V$ is defined by the condition
\begin{equation}
\hat{g}_{ij}|\mu\rangle\equiv0\ \ \forall\ i<j,
\label{eq:hws}
\end{equation}
i.e. the highest weight state vanishes under the operation of any raising operator. 
Note that $|\mu\rangle$ is also the common eigenvector of the Cartan subalgebra generators
\begin{equation}
\hat{H}_{1}|\mu\rangle=\frac{1}{2}\lambda_{1}|\mu\rangle,\ldots,\hat{H}_{N-1}|\mu\rangle=\frac{1}{2}\lambda_{N-1}|\mu\rangle.
\end{equation}
The $N-1$ eigenvalues $[\lambda_{1},\ldots,\lambda_{N-1}]$ are used to label the irreps of SU($N$) and the dimension of the representation is given by  Weyl's dimension formula~\citep{cornwell1997group}.
For the sake of simplicity, we will focus on $\mathfrak{su}(2)$ and $\mathfrak{su}(3)$. The Cartan subalgebra of $\mathfrak{su}(2)$ is one-dimensional, and therefore its representation is labeled by a single number $\lambda_{1}$ with $\dim[\lambda_{1}]=\lambda_{1}+1$, c.f. the familiar result $\dim[S]=2S+1$ can be recovered by setting $S=\lambda_{1}/2$. 
Whereas for $\mathfrak{su}(3)$, the irreducible representations are labeled by $\lambda_1$ and $\lambda_2$ and the dimension of the $[\lambda_{1},\lambda_{2}]$ is given by $\text{dim}[\lambda_{1},\lambda_{2}]=\frac{1}{2}(\lambda_{1}+1)(\lambda_{2}+1)(\lambda_{1}+\lambda_{2}+2).$
For example, for a spin one system (three-level system) with a basis of states $\{|S^z=1\rangle, |S^z=-1\rangle, |S^z=0\rangle \}$, the generators of the Cartan subalgebra are ${\hat H}_1={\hat S}^z/2$ and ${\hat H}_2=3({\hat S}^z)^2/4-{\hat S^z}/4-1/2$ and the highest-weight state of the above-mentioned fundamental representation of $\mathfrak{su}(3)$ is the state $|1\rangle$ (${\hat S}^z |1\rangle = |1\rangle$) with $\lambda_{1}=1$ and $\lambda_{2}=0$, implying that the fundamental representation  has $\text{dim}[1,0]=3$. Note that the fundamental
representation is a special example of a so-called ``\emph{degenerate representation}'' with only one non-zero eigenvalue $\lambda_{k}$.
In this paper we will focus on the particular degenerate representations $[\lambda_1, 0, \ldots, 0]$ that have a non-zero eigenvalue only for $\hat{H}_1$. 

\subsection{SU($N$) coherent states}

In this section we discuss the construction of SU($N$) coherent states for degenerate representations. We start by considering the simplest non-trivial case corresponding to  coherent states of SU(2). The three operators 
\begin{equation}
\hat{S}^{+}=\hat{g}_{12},\ 
\hat{S}^{-}=\hat{g}_{21},\
\hat{S}^{z}=\hat{H}_{1},
\label{eq:su2_nohbar}
\end{equation}
form a basis of $\mathfrak{su}(2)$ generators.
The corresponding highest-weight state $|\mu\rangle=|S^z=S\rangle$ (hereafter we will use $\hbar$ as the unit of angular momentum), which satisfies $\hat{S}^{+}|S^z=S\rangle=0$, is chosen as the reference state.
Like any other state, $|\mu\rangle$ is defined up to an arbitrary multiplicative phase. To remove this redundancy in the definition of coherent states, it is necessary to identify the \emph{isotropic subgroup} $I$, that leaves the reference state invariant up to a multiplicative  phase~\citep{ZhangRMP}. Since  $I=\text{U(1)}$ for the particular case of SU(2),  the manifold of coherent states is isomorphic to the coset space $\hat{\Omega}\in\text{SU}(2)/\text{U}(1)$:
\begin{equation}
|\Omega(\theta,\phi)\rangle\equiv\hat{\Omega}(\theta,\phi)|S^{z}=S\rangle=e^{-i\hat{S}^{z}\phi}e^{-i\hat{S}^{y}\theta}|S^{z}=S\rangle,
\end{equation}
where $\hat{S}^{y}=\hat{S}^+-i\hat{S}^-$ and $\theta,\phi$ are two real parameters that parametrize the two-sphere $S^{2}\simeq \text{CP}^{1}$. In general, the manifold of the coherent states is isomorphic to $G/I$~\citep{Gnutzmann_1998,ZhangRMP}, and it is known as the \emph{quotient orbit}. In the fundamental representation of SU(2), an arbitrary SU(2) coherent state can be expressed as
\begin{equation}
|\Omega(\theta,\phi)\rangle=\cos\frac{\theta}{2}e^{-i\phi/2}|1\rangle+\sin\frac{\theta}{2}e^{i\phi/2}|2\rangle.
\label{eq:SU2CS}
\end{equation}
Let us consider now the case of SU(3) coherent states. The highest-weight state is again chosen to be the reference state. To identify the isotropic subgroup for degenerate representations, without loss of generality, we consider the highest-weight state $|1\rangle=(1,\ 0,\ 0)^T$ in the fundamental representation of SU(3). Note that  this state is  invariant  under the SU(2) group of transformations restricted to the orthogonal subspace. In addition, the global multiplication by a phase [U(1) subgroup] also leaves the reference state invariant in the quantum mechanical sense, implying that the isotropic group under consideration is $I = \text{SU}(2) \times \text{U}(1) \simeq \text{U}(2)$. The resulting manifold of SU(3) coherent states is then isomorphic to the coset space $\text{SU}(3)/ \text{U}(2)\simeq S^{5}/S^{1}\simeq \text{CP}^{2}$.
Note that the dimension of this manifold is $8 - 4= 4$.
In the fundamental representation, a generic SU(3) coherent state can be expressed as
\begin{multline}
|\Omega(\theta,\phi,\alpha_{1},\alpha_{2}) \rangle 
   = R(\theta,\phi,\alpha_{1},\alpha_{2})| 1 \rangle \\
   = e^{i\alpha_{1}}\sin\theta\cos\phi|1\rangle+e^{i\alpha_{2}}\sin\theta\sin\phi|2\rangle + \cos\theta|3\rangle,
\label{eq:SU3CS}
\end{multline}
where $R(\theta,\phi,\alpha_{1},\alpha_{2})\in\text{SU}(3)/\text{U}(2)$, that takes the form~\citep{Sen_SU3}
\begin{equation}
\begin{pmatrix}
\sin\theta\cos\phi e^{i\alpha_{1}} & \cos\theta\cos\phi e^{i\alpha_{1}} & -\sin\phi e^{-i\alpha_{2}}\\
\sin\theta\sin\phi e^{i\alpha_{2}} & \cos\theta\sin\phi e^{i\alpha_{2}} & \cos\phi e^{-i\alpha_{1}}\\
\cos\theta & -\sin\theta & 0
\end{pmatrix}.
\end{equation}

A general SU($N$) coherent state with $N>3$ can be constructed by a straightforward generalization of the procedure that we used for the SU(2) and SU(3) cases:
\begin{equation}
|\Omega(\{p_i\})\rangle\equiv\hat{\Omega}(\{p_i\})|\mu\rangle,
\end{equation}
where $\hat{\Omega}(\{p_i\})\in \text{SU}(N)/I$ with $I \simeq \text{SU}(N-1) \times \text{U}(1) \simeq \text{U}(N-1)$. The SU($N$) coherent state, which is topologically equivalent to the complex projective space CP$^{N-1}$, is then parameterized by  $N^2-1-(N-1)^2=2(N-1)$ real parameters $\{p_i\}$. The explicit forms of the SU($N$) coherent states can be found in Ref.~\citep{Nemoto_2000} [in the spherical coordinates with $2(N-1)$ real parameters] and in Ref.~\citep{Gitman_1993} (in terms of $N-1$ complex parameters).
Similarly to the $H_{3}$ coherent states  [see Eq.~\eqref{eq:overlap}], the SU($N$) coherent states form an over-complete basis of the Hilbert space of an $N$-level system. The overlap between two SU($N$) coherent states and the corresponding integration measure over the parameters $\{p_i\}$ can be found in Ref.~\citep{Nemoto_2000}.


\section{Classical limit for a spin system based on SU($N$) coherent states\label{sec:SUN_classical}}

We have seen in Sec.~\ref{sec:point1D}  that the $H_3$ coherent states can be used to recover the classical mechanics ($\hbar\rightarrow0$ limit) of a point particle.
In this section we will follow  similar steps to obtain a classical limit of a quantum spin-$S$ system, whose local Hilbert space has a dimension $N=2S+1$~\cite{Gitman_1993}. Since the SU($N$) group consists of all unitary basis transformations (with determinant equal to 1) of an $N$-level quantum-mechanical system, coherent states of SU($N$) provide a natural platform to define the classical limit of a spin system. However, as we mentioned previously, the traditional approach has always been to use SU(2) coherent states to define a classical limit of spin systems. While the SU($N$) Lie group is not the only alternative, the purpose of this section is to introduce this alternative path towards a classical a limit and later illustrate the advantages of using Lie groups that are larger than SU(2). 

As before, the first step of the process is to take the $\hbar\rightarrow0$ limit of the expectation value of a quantum operator in an SU($N$) coherent state. We then start by considering as an example the classical limit of the \emph{physical spin operator} $\hat{S}^{z}=\hbar\hat{H}_{1}$ in the highest-weight state of SU(2):
\begin{equation}
    \lim_{\hbar\rightarrow0} \langle S^z=S | \hat{S}^{z} | S^z=S \rangle = \lim_{\hbar\rightarrow0} \hbar S.
    \label{eq:szclassical}
\end{equation}
Note that we made $\hbar$ explicit in the above equation to indicate that the spin is an intrinsically quantum-mechanical object, i.e., the simple definition of the classical operator given in Eq.~\eqref{eq:expcl} does not exist for $\hat{S}^z$ in Eq.~\eqref{eq:szclassical}. However, we can still obtain a non-trivial classical limit by simultaneously sending $S=\lambda_1/2$ to infinity and $\hbar$ to zero, while keeping the product finite. This simple procedure can be easily generalized to any quantum operator $\hat{A}$, which is a polynomial function of the physical SU($N$) generators. The classical limit of an  operator is then defined by taking the expectation value on a coherent state  
\begin{equation}
a(\{\alpha_{\rho}\})=\langle\Omega(\{\alpha_{\rho}\})|\hat{A}|\Omega(\{\alpha_{\rho}\})\rangle,\label{eq:gclassical}
\end{equation}
where $\{\alpha_{\rho}\}$ is the set of $N-1$ complex parameters that parametrize the SU($N$) coherent states, and simultaneously sending the eigenvalue $\lambda_1$ --degenerate representations of SU($N$)-- to infinity.

The second step is to consider the classical limit of the expectation value of the product of two operators.
We ought to prove that the factorization rule holds in the classical limit
\begin{equation}
\langle\Omega(\{\alpha_{\rho}\})|\hat{A}\hat{B}|\Omega(\{\alpha_{\rho}\})\rangle\xrightarrow{\text{classical limit}}a(\{\alpha_{\rho}\})b(\{\alpha_{\rho}\}).\label{eq:factorization}
\end{equation}
Indeed, the factorization rule holds for $\lambda_1\rightarrow\infty$. The explicit proof is provided in Appendix~\ref{apd:su2} for the SU(2) case. The general proof for SU($N$) is non-trivial and can be found in Ref.~\citep{Gitman_1993}.  The crucial observation is that two distinct SU($N$) coherent states become orthogonal in the limit $\lambda_1\rightarrow\infty$. As a result, one can expand the integral associated with the expectation value in a similar manner as in Eq.~\eqref{eq:abexpand}, but now in powers of $1/\lambda_1$. Finally, by expanding the expectation value of the commutator between two operators up to the first order in $1/\lambda_1$ and then sending $\lambda_1\rightarrow\infty$, one can prove the generalization of Dirac's conjecture applied to the orbit of the SU($N$) coherent states
\begin{multline}
\{a(\{\alpha_{\rho}\}),b(\{\alpha_{\rho}\})\}_{PB} 
=\sum_{\mu,\nu}g^{\mu\nu}\bigg(\frac{\partial a}{\partial \alpha_{\nu}}\frac{\partial b}{\partial \bar{\alpha}_{\mu}}-\frac{\partial a}{\partial \bar{\alpha}_{\mu}}\frac{\partial b}{\partial \alpha_{\nu}}\bigg)  \\
=\lim_{\lambda_1\rightarrow\infty}-i\lambda_1\langle\Omega(\{\alpha_{\rho}\})|[\hat{A},\hat{B}]|\Omega(\{\alpha_{\rho}\})\rangle,
\label{eq:dirac2}
\end{multline}
where $g^{\mu\nu}$ is the  Fubini-Study metric of CP$^{N-1}$~\citep{helgason2001differential}. The Poisson bracket of SU(2) is derived in Appendix~\ref{apd:su2} and the general derivation for SU($N$) is given in Ref.~\citep{Gitman_1993}. We want to point out that for practical purposes, it is not always necessary to know \emph{a priori} the exact form of the Poisson bracket to obtain the classical equations of motion. In most cases, it is simpler to evaluate the right hand side of Eq.~\eqref{eq:dirac2}. Namely, Eq.~\eqref{eq:dirac2} provides us with a simple recipe to derive the classical dynamics on the manifold of SU($N$) coherent states. We will follow this recipe in the next section to write down the generalized classical equations of motion for a quantum spin system.

\section{Classical equations of motion for spins\label{sec:classical_eom}}

In the previous section we saw that the classical SU($N$) dynamics of a quantum spin system becomes exact for $\lambda_1\rightarrow\infty$.
However, for most systems of interest, the dimension of the local Hilbert space is finite and $\lambda_1=1$, implying that the classical dynamics is just an approximation of the exact quantum dynamics.
The big advantage of this approximation is that the numerical cost of the simulations drops from an exponential to a linear dependence in the number of spins. 
The mathematical procedure of taking the classical limit can be physically represented as  building a large SU($N$) spin by ferromagnetically coupling $M$ replicas of the original spin  and then sending $M$ to infinity. By replacing the SU($N$) representation $[\lambda_1,\ldots,0]$ with $[M\lambda_1,\ldots,0]$, the spin---originally a microscopic object---becomes a macroscopic entity. It is important to note that the ``ferromagnetic'' coupling between replicas corresponds to an SU($N$) ferromagnetic Heisenberg interaction~\cite{auerbach2012interacting}. In other words, the difference between different classical limits of a given spin system [e.g. SU(2) and SU($N$)] is dictated by the \emph{nature of the coupling} between different replicas. 
This raises the question about which classical limit better approximates the exact quantum dynamics. As we will explain in this section, the short answer to this question is that the choice of SU($N$) coherent states with $N=2S+1$ guarantees that the dynamics will capture all the coherent low-energy modes that can appear for a general Hamiltonian. To appreciate this important point, we first derive two \emph{different} classical dynamics based on SU(2) and SU(3) coherent states for a \emph{common} quantum spin Hamiltonian and  then we provide the general recipe to derive the classical dynamics for SU($N$) coherent states. 

\subsection{SU(2) and SU(3) Landau-Lifshitz dynamics\label{subsec:classical_su2su3}}

Consider the following $S=1$ spin Hamiltonian
\begin{equation}
\mathcal{\hat{H}}=\frac{1}{2}\sum_{\bm{r},\bm{\delta}}\sum_{\alpha,\beta}\hat{S}_{\bm{r}}^{\alpha}J_{\bm{\delta}}^{\alpha\beta}\hat{S}_{\bm{r}+\bm{\delta}}^{\beta}+D\sum_{\bm{r}}(\hat{S}_{\bm{r}}^{z})^{2},\label{eq:model}
\end{equation}
where $J_{\bm{\delta}}^{\alpha\beta}$ is the exchange tensor on the bond $\bm{\delta}$ and $D$ is the strength of a single-ion anisotropy term. This $S=1$ system admits two classical limits: one based on SU(2) coherent states (second representation with $\lambda_1=2$) and another one based on the SU(3) coherent states (fundamental representation with $\lambda_1=1$ and $\lambda_2=0$). 

We will first consider the SU(2) classical limit. The time evolution of the spin components [generators of SU(2)] is dictated by the Heisenberg equation of motion
\begin{align}
\frac{d\hat{S}_{\bm{r}}^{\alpha}}{dt} & =-\frac{i}{\hbar}\big[\hat{S}_{\bm{r}}^{\alpha},\hat{\mathcal{H}}\big]\nonumber \\
 & =\sum_{\bm{\delta}}\sum_{\mu,\nu,\beta}J_{\bm{\delta}}^{\mu\nu}\epsilon^{\alpha\mu\beta}\hat{S}_{\bm{r}}^{\beta}\hat{S}_{\bm{r}+\bm{\delta}}^{\nu}\nonumber \\
 & +D\sum_{\beta}\epsilon^{\alpha3\beta}(\hat{S}_{\bm{r}}^{\beta}\hat{S}_{\bm{r}}^{z}+\hat{S}_{\bm{r}}^{z}\hat{S}_{\bm{r}}^{\beta}),\label{eq:spineom}
\end{align}
where $\epsilon^{\alpha\mu\beta}$ is the Levi-Civita symbol. The classical limit of the interaction term is given by
\begin{multline}
\hat{S}_{\bm{r}}^{\beta}\hat{S}_{\bm{r}+\bm{\delta}}^{\nu} \rightarrow \langle\Omega|\hat{S}_{\bm{r}}^{\beta}\hat{S}_{\bm{r}+\bm{\delta}}^{\nu}|\Omega\rangle \\ =\langle\Omega_{\bm{r}}|\hat{S}_{\bm{r}}^{\beta}|\Omega_{\bm{r}}\rangle\langle\Omega_{\bm{r}+\bm{\delta}}|\hat{S}_{\bm{r}+\bm{\delta}}^{\nu}|\Omega_{\bm{r}+\bm{\delta}}\rangle
=s_{\bm{r}}^{\beta}s_{\bm{r}+\bm{\delta}}^{\nu}\label{eq:sionfact}
\end{multline}
where we assumed that the coherent state of this system is a direct product of local coherent states, i.e., $|\Omega\rangle=\otimes_{\bm{r}}|\Omega_{\bm{r}}\rangle$. 
Since the single-ion anisotropy term is quadratic in the spin operators, we must use the factorization rule given in Eq.~\eqref{eq:factorization}, which is only exact in the  classical limit
\begin{equation}
 \hat{S}_{\bm{r}}^{\beta}\hat{S}_{\bm{r}}^{z}+\hat{S}_{\bm{r}}^{z}\hat{S}_{\bm{r}}^{\beta}
 \rightarrow \langle\Omega_{\bm{r}}|\hat{S}_{\bm{r}}^{\beta}\hat{S}_{\bm{r}}^{z}+\hat{S}_{\bm{r}}^{z}\hat{S}_{\bm{r}}^{\beta}|\Omega_{\bm{r}}\rangle
 \xrightarrow{\lambda_1 \rightarrow\infty} 2s_{\bm{r}}^{\beta}s_{\bm{r}}^{z},
\end{equation}
As for the left-hand-side of the HE, the classical limit simply gives
\begin{equation}
d\hat{S}_{\bm{r}}^{\alpha}/dt\rightarrow \langle\Omega_{\bm{r}}|d\hat{S}_{\bm{r}}^{\alpha}/dt|\Omega_{\bm{r}}\rangle
=ds_{\bm{r}}^{\alpha}/dt.
\end{equation}
Consequently, the resulting  classical equation of motion for  SU(2) coherent states takes the form
\begin{equation}
\frac{ds_{\bm{r}}^{\alpha}}{dt}=\sum_{\bm{\delta}}\sum_{\mu,\nu,\beta}J_{\bm{\delta}}^{\mu\nu}\epsilon^{\alpha\mu\beta}s_{\bm{r}}^{\beta}s_{\bm{r}+\bm{\delta}}^{\nu}+2D\epsilon^{\alpha3\beta}s_{\bm{r}}^{\beta}s_{\bm{r}}^{z},
\label{eq:su2ll}
\end{equation}
which is the well-known Landau-Lifshitz (LL) equation without the damping term,
\begin{equation}
    \frac{d s_{\bm{r}}^{\alpha}}{d t} = \sum_{\mu \nu} \epsilon_{\alpha \mu \nu} {s}^{\mu}_{\bm{r}}  {b}^{\nu}_{\bm{r}}, \quad \bm{b}_{\bm{r}}=-\frac{d h}{d \bm{s}_{\bm{r}}},
    \label{eq:tradll}
\end{equation}
where $h=\langle \Omega|\hat{\mathcal{H}}| \Omega \rangle$ is the classical Hamiltonian.

Let us consider now the classical limit based on  SU(3) coherent states. In this case we need to compute  the equation of motion of the eight  generators of SU(3), $\hat{O}_{1-8}$, which can be regarded as the components of the SU(3) spin.
For this purpose, we are going to use the basis of generators given in Eq.~\eqref{eq:su3_physicalop}, which are  obtained by applying an SU(3) transformation (change of basis) to the standard basis  presented in Eq.~\eqref{eq:sun_basis}. The advantage of the new ``physical'' basis is that it is  a direct sum of bases of irreps of the SO(3) group of rotations. The first three elements, $\hat{O}^{1-3}$, are the three spin operators $\hat{S}^{\alpha}$ which represent  a local dipole moment and transform as vectors under rotations [three-dimensional  irrep of SO(3)]. The last five elements of the basis, $\hat{O}^{4-8}$, are symmetric and traceless bilinear forms in the spin operators that represent a nematic or quadrupolar moment [five-dimensional irrep of SO(3)]. Each irrep of SO(3) corresponds to a different multipole and the number of different irreps or multipoles for the more general SU($N$) group is $N-1$. Importantly, the equations of motion that dictate the dynamics of the different multipolar components are coupled. For instance,
as we will see below, the dynamics of the dipolar generators of SU(3), $\hat{O}^{1-3}$, is coupled to the dynamics of the nematic generators $\hat{O}^{4-8}$.
By adopting the SU(3) approach, we are treating the dipolar and the quadrupolar components on equal footing. 

The first step is to rewrite the Hamiltonian  in terms of the SU(3) generators:
\begin{equation}
    \mathcal{\hat{H}} = \frac{1}{2}\sum_{\bm{r},\bm{\delta}}\sum_{i,j=1}^3\hat{O}_{\bm{r}}^{i}J_{\bm{\delta}}^{ij}\hat{O}_{\bm{r}+\bm{\delta}}^{j}+\frac{D}{\sqrt{3}}\sum_{\bm{r}}\left(\hat{O}_{\bm{r}}^{8} + \frac{2}{\sqrt{3}}\right).
\end{equation}
Note that the second term can be expressed in multiple ways. This ambiguity is removed by requiring that each Hamiltonian term \emph{must be linear in the SU(3) generators} acting on a given site ${\bm{r}}$ [note that the generators of SU($N$) together with the identity form a complete basis for the complex vector space of $N\times N$ matrices].
As shown in Eq.~\eqref{eq:su3_commu}, the quadrupolar components $\hat{O}^{4-8}$ in general do not commute with the dipolar ones $\hat{O}^{1-3}$. As a result, the eight Heisenberg equations of motion for the SU(3) spin components on each site turn out to be coupled:
\begin{equation}
    \frac{d \hat{O}_{\bm{r}}^i}{d t} 
    = \sum_{\bm{\delta}} \sum_{j,k=1}^3\sum_{l=1}^8 J_{\bm{r}}^{jk} f_{ijl} \hat{O}_{\bm{r}}^l \hat{O}_{\bm{r}+\bm{\delta}}^{k} 
    + \frac{D}{\sqrt{3}} \sum_{l=1}^8 f_{i8l} \hat{O}_{\bm{r}}^l.
\label{su3He}
\end{equation}
Here $i=1\ldots8$ and $f_{ijl}$ are the structure constants of $\mathfrak{su}(3)$ in the physical basis [see Eq.~\eqref{eq:su3_strconst} for the non-zero structure constants]. As with the SU(2) case, the classical limit of the interaction term is again obtained by assuming that the coherent state is a direct product of coherent states on each site. An important difference, however, is that the classical limit of the single-ion term does not require the use of the factorization rule, which can be a strong approximation for finite values of $\lambda_1$, because each term of 
Eq.~\eqref{su3He} is now linear in the generators of SU(3) acting on a given site ${\bm{r}}$. 
After taking the classical limit on the left-hand side of the above equation, we obtain the classical equations of motion for the SU(3) spins
\begin{equation}
    \frac{d o_{\bm{r}}^i}{d t} 
    = \sum_{\bm{\delta}} \sum_{j,k=1}^3\sum_{l=1}^8 J_{\bm{r}}^{jk} f_{ijl} o_{\bm{r}}^l o_{\bm{r}+\bm{\delta}}^{k} 
    + \frac{D}{\sqrt{3}} \sum_{l=1}^8 f_{i8l} o_{\bm{r}}^l,
\label{su3LL}
\end{equation}
or
\begin{equation}
    \frac{d o_{\bm{r}}^{\alpha}}{d t} = \sum_{\mu \nu} f_{\alpha \mu \nu} {o}^{\mu}_{\bm{r}}  {b}^{\nu}_{\bm{r}}, \quad \bm{b}_{\bm{r}}=-\frac{d h}{d \bm{o}_{\bm{r}}}.
    \label{eq:su3LL2}
\end{equation}
Eq.~\eqref{eq:su3LL2} can be regarded as a generalization of the LLD \eqref{eq:tradll}. 

Clearly,  the SU(3) approach becomes strictly necessary when the ground state  of the Hamiltonian under consideration has some form of nematic ordering, $\langle \hat{O}^{1-3} \rangle=0$ and $\langle \hat{O}^{\nu} \rangle \neq 0$ for some values of $4 \leq \nu \leq 8$, which can either be spontaneous or induced by a large single-ion anisotropy term, such as the last term of $\mathcal{\hat{H}}$ for  $D \gg |J_{\bm{\delta}}^{\alpha\beta}|$. The simple reason is that the SU(2) coherent states cannot describe a local quadrupolar moment. The need for the SU(3) dynamics becomes a bit more subtle when the ground state exhibits some form of magnetic (dipolar) ordering. Even in that case, nematic fluctuations can renormalize the magnitude of the dipole moment or produce coherent low-energy modes, which are different from the usual spin waves (dipolar fluctuations). These are the situations in which the SU(3) dynamics becomes more appropriate than the traditional SU(2) dynamics. In a few words, 
the SU(3) approach can faithfully represent all types of local fluctuations of a  three-level system. 




\subsection{SU($N$) Landau-Lifshitz dynamics}

The SU(2) LLD can be straightforwardly generalized to SU($N$) spins by following the same steps that we described for the SU(3) case. The classical equations of motion are obtained by (i) expressing the Hamiltonian in terms of generators of SU($N$) under the condition that each term must be linear in the SU($N$) spin components acting on a given site, (ii) computing the Heisenberg equation of motion for each generator of SU($N$), and (iii) replacing the operators with their expectation values for coherent states of SU($N$).

Based on the above discussions, it is apparent that to faithfully describe all types fluctuations in a spin-$S$ system, whose classical phase space is isomorphic to CP$^{N-1}$, we need to take the classical limit based on SU($N$) coherent states with $N=2S+1$. We note however that for particular spin Hamiltonians a subgroup of SU($N$) may also provide a good approximation, as long as it incorporates the relevant components of the local order parameter. As an example, we can just consider the case of pure isotropic Heisenberg models, whose dynamics is well described by the traditional SU(2) LLD if the spin $S$ is large enough. The key observation is that the local order parameter of these systems has a dominant dipolar character and the normal modes are coherent spin waves. Non-dipolar fluctuations can still be described as a continuum of multiple spin waves. In other words, the SU(2) dynamics breaks down when new normal modes associated with non-dipolar fluctuations emerge below the continuum of multiple spin waves. 


\begin{figure*}[ht!]
    \centering
    \includegraphics[width=1.0\textwidth]{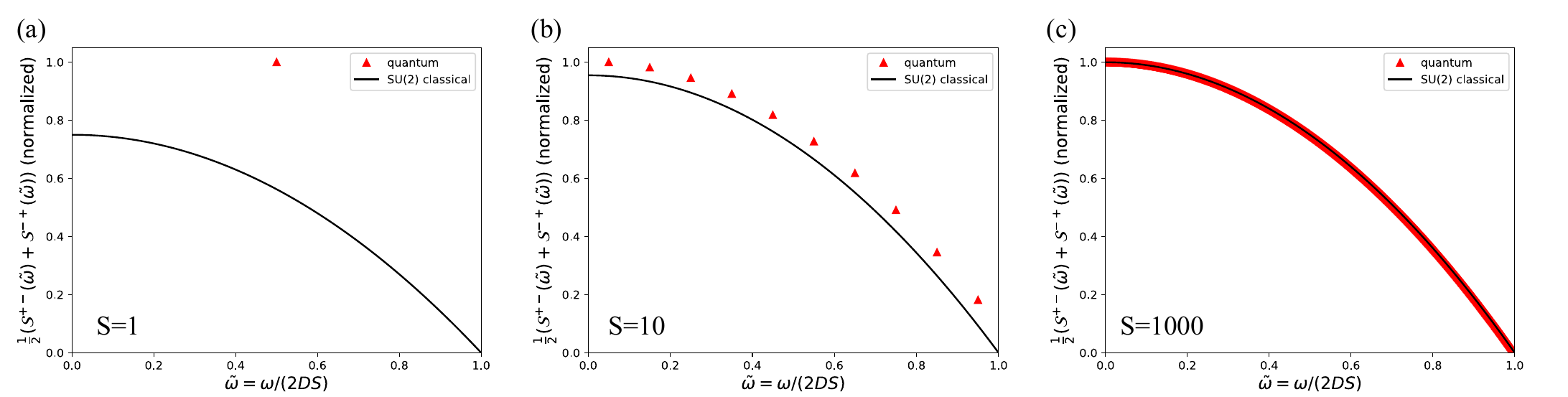}
    \caption{Comparisons of the transverse dynamical spin structure factor between the quantum (red triangle) and the SU(2) classical results (black line) for three different spin values: (a) $S=1$, (b) $S=10$, and (c) $S=1000$.
    In all three panels, the inverse temperature $\beta DS^2=10^{-5}$, and the values of the dynamical
    structure factor are normalized to  the maximum intensity of the exact result.}
    \label{fig:fig1}
\end{figure*}
\begin{figure}[ht!]
    \centering
    \includegraphics[width=.45\textwidth]{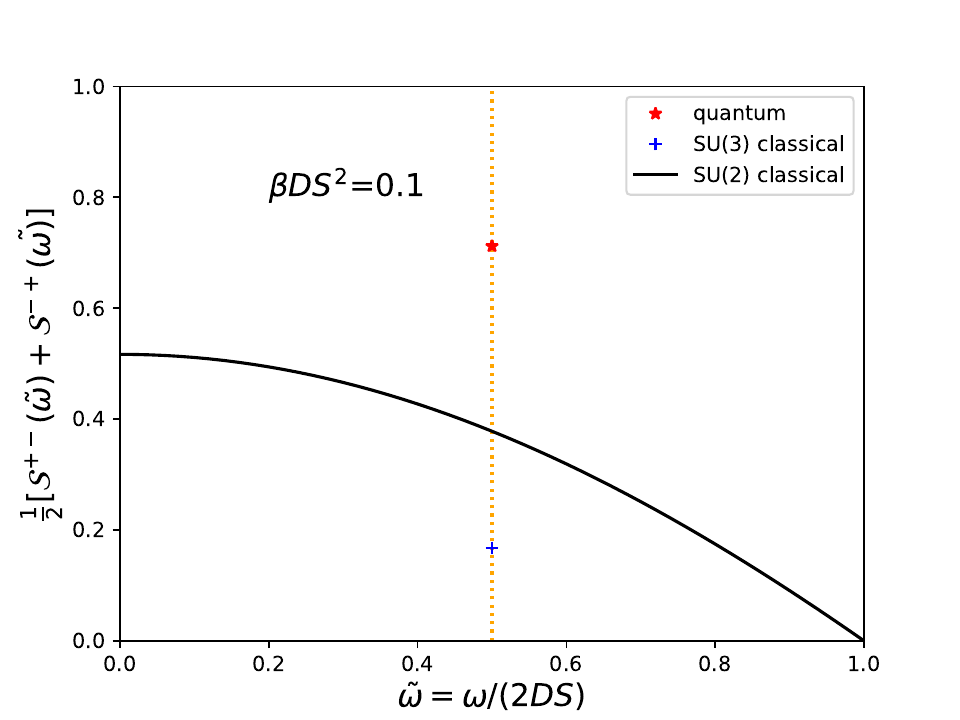}
    \caption{Comparisons of the transverse dynamical spin structure factor for the $S=1$ single-ion model as a function of the energy transfer at the inverse temperature $\beta D S^2=0.1$: quantum (red star), SU(3) classical (blue cross), and SU(2) classical
    (black line).}
    \label{fig:fig2}
\end{figure}

\section{Single-ion model\label{sec:sion}}

Consider the following integer spin-$S$ single-ion Hamiltonian,
\begin{equation}
\hat{\mathcal{H}}_{\rm SI}=D(\hat{S}^{z})^{2},
\label{eq:sion}
\end{equation}
with $D>0$. This term controls the high-temperature dynamics of the Hamiltonian given in Eq.~\eqref{eq:model} if  $|D| \gg |J^{ij}_{\bm{\delta}}|$. The non-degenerate ground state of $\hat{\mathcal{H}}_{\rm SI}$ is the eigenstate of $S^{z}$ with eigenvalue $m=0$: $|m=0\rangle$. The excited states are the doublets $|\pm m\rangle$ with $1\leq m\leq S$. The exact quantum dynamics can be solved in a closed form as there is no interaction term in the Hamiltonian. The transverse dynamical spin structure factor can be computed by using the Lehmann representation:
\begin{align}
\mathcal{S}^{+-}(\omega) & = \frac{1}{2\pi Z} \int_{-\infty}^{+\infty} {\rm d}t 
                  e^{i\omega t} \Tr\left[e^{-\beta\hat{\mathcal{H}}_{\text{SI}}} \hat{S}^{+}(t)\hat{S}^{-}(0) \right] \nonumber\\
               & = \frac{1}{Z} \sum_{m=0}^{S-1}
                  e^{-\beta m^{2}D}(S-m)(S+m+1) \delta[\omega-(2m+1)D] \nonumber \\
               & =\mathcal{S}^{-+}(\omega),
\label{eq:spm_qm}
\end{align}
for $\omega\geq0$, where
\begin{equation}
Z=1+2\sum_{m=1}^{S}e^{-\beta m^{2}D}
\end{equation}
is the partition function. Similarly,
\begin{align}
\mathcal{S}^{zz}(\omega)  = \frac{1}{2\pi Z} \!\! \int_{-\infty}^{+\infty} \!\!\!\!\!\!\!\!\!\! {\rm d}t
                  e^{i\omega t} \Tr\left[e^{-\beta\hat{\mathcal{H}}_{\text{SI}}} \hat{S}^{z}(t)\hat{S}^{z}(0) \right]
                = -\frac{1}{\beta} \frac{\partial Z}{\partial D} \delta(\omega).
\end{align}

The classical spin dynamics of this problem can also be solved analytically. Let us first consider the  SU(2) LL equations:
\begin{equation}
\frac{ds^{x}}{dt}=-2Ds^{y}s^{z},\quad \frac{ds^{y}}{dt}=2Ds^{x}s^{z},\quad \frac{ds^{z}}{dt}=0.
\end{equation}
The solution is given by
\begin{equation}
s^{+}(t)=s^{+}(0)e^{i\tilde{\omega}_{-}t},\quad s^{-}(t)=s^{-}(0)e^{i\tilde{\omega}_{+}t},\quad
s^{z}(t)=s^{z}(0),
\end{equation}
with $\tilde{\omega}_{\pm}=\pm 2Ds_{z}(0)$. The corresponding transverse dynamical spin structure factor is computed by performing a thermal average over initial  SU(2) coherent states (see Appendix.~\ref{apd:corrcl} for details), which reads
\begin{align}
    \mathcal{S}^{+-}_{\text{SU(2)}}(\omega) 
    & = \frac{S^{2}\sqrt{\beta D} \left(1-\frac{\omega^{2}}{4D^{2}S^{2}}\right)
                   e^{-\beta\frac{\omega^{2}}{4D}}}
                   {2D\sqrt{\pi}\text{erf}(\sqrt{\beta D}S)}\Theta(2DS-|\omega|) \nonumber\\
    & =\mathcal{S}^{-+}_{\text{SU(2)}}(\omega),
    \label{eq:spm_su2}
\end{align}
where $\text{erf}(x)$ is the error function and $\Theta(x)$ is the Heaviside step function. Similarly,
\begin{equation}
    \mathcal{S}^{zz}_{\text{SU(2)}} = \delta(\omega)S^{2}\bigg[-\frac{e^{-\alpha}}{\sqrt{\alpha}}
                   \frac{1}{\sqrt{\pi}\text{erf}(\sqrt{\alpha})}+\frac{1}{2\alpha}\bigg],
\end{equation}
where $\alpha=\beta DS^{2}$.

Figure~\eqref{fig:fig1} shows a comparison between the exact result for the transverse dynamical spin structure factor [see Eq.~\eqref{eq:spm_qm}] and the SU(2) classical approximation given in Eq.~\eqref{eq:spm_su2} for three different spin values. Clearly, the quantum mechanical spectrum consists of discrete absorption peaks corresponding to $\Delta S_z = 1$ (for $\omega>0$) transitions between discrete energy levels, whereas  the SU(2) classical theory produces a broad continuum of spectral weight centered around $\omega=0$. The continuous character of the distribution arises from the dependence of the  spin precession frequency on the conserved quantity, $s^z(t)=s^z(0)$, which has a continuous distribution on the orbit of the SU(2) coherent states. For small values of $S$, such as $S=1$, the classical result based on SU(2) coherent states deviates strongly from the exact quantum mechanical result [see Figs.~\eqref{fig:fig1}(a) and 1(b)]. Only for very large  values of $S$ [see Fig.~\eqref{fig:fig1} (c) for $S=1000$]  the SU(2) classical result becomes a good approximation. The slow convergence of the quantum mechanical result to the large-$S$ limit  demonstrates the need of implementing an alternative classical limit for  realistic Hamiltonians, such as $\hat{\mathcal{H}}_{\rm SI}$.


To illustrate the ideas discussed in Sec.~\ref{subsec:classical_su2su3}, we now consider the classical limit of the single-ion model based on SU(3) coherent states as an approximation for the extreme $S=1$ case. The generalized SU(3) LL equations take the form
\begin{equation}
\frac{do_{1}}{dt}=Do_{5},\quad
\frac{do_{2}}{dt}=-Do_{4},\quad 
\frac{do_{4}}{dt}=Do_{2},\quad
\frac{do_{5}}{dt}=-Do_{1},\quad
\end{equation}
\begin{equation}
\frac{do_{3}}{dt} = \frac{do_{6}}{dt} = 
\frac{do_{7}}{dt} = \frac{do_{8}}{dt} = 0.
\end{equation}
The solution gives the time evolution of the classical dipole operators
\begin{align}
s^{\pm}(t) & = \frac{1}{2} \big[ s^{\pm}(0) \mp o_{4}(0) - io_{5}(0) \big]e^{iDt} \nonumber\\
           & + \frac{1}{2} \big[ s^{\pm}(0) \pm o_{4}(0) + io_{5}(0) \big]e^{-iDt},
\end{align}
where $s^{\pm}=o_{1}\pm io_{2}$, and
\begin{equation}
s^{z}(t)=s^{z}(0)=o_{3}(0)
\end{equation}
is a conserved quantity. By computing the thermal average over initial SU(3) coherent states of the Fourier transform of the spin-spin correlation function, we obtain the dynamical spin structure factor for $\omega\geq0$:
\begin{align}
    {\cal S}^{+-}_{\text{SU(3)}}(\omega) & = \delta_{\omega, D} 
    \frac{6+2e^{\beta D} (\beta D-3) + \beta D(\beta D +4)}
    {\beta^2 D^2 (e^{\beta D}-\beta D -1)} \nonumber \\
    & = S^{-+}_{\text{SU(3)}}(\omega),
    \label{eq:spm_su3}
\end{align}
and
\begin{equation}
{\cal S}^{zz}_{\text{SU(3)}}(\omega)=\delta_{\omega, 0}\frac{6e^{\beta D} - \beta D \left[ 6+ \beta D\left(3+\beta D\right)\right]-6}
{3\beta^2 D^2 (e^{\beta D}-\beta D-1)}.
\end{equation}

Figure~\eqref{fig:fig2} shows the comparison between the exact transverse dynamical spin structure factor of the $S=1$ single-ion model and the results obtained with the  SU(2) and SU(3) classical approximations for $\beta DS^2=0.1$. 
We can see that by working with the CP$^{2}$ classical phase space of the $S=1$ system, we obtain a single transition at the correct frequency $\omega = D$, as opposed to the SU(2) classical approach. We also note that in the infinite temperature limit, $\beta DS^2 \to 0$, the intensity of the SU(3) classical result coincides with the exact quantum mechanical result if we renormalize the classical SU(3) spins $O^j_{\bm{r}}  \to  \kappa  O^j_{\bm{r}} $ with $\kappa(N=3,\lambda_1=1)=2$.
This renormalization factor is also required to satisfy the sum rule: 
\begin{equation}
\sum_{j} \int d\omega  {\cal O}^{jj}(\omega) = C_1(N=3,\lambda_1=1) = \frac{16}{3},
\end{equation}
associated with the quadratic Casimir operator of SU(3) with eigenvalue $C_1(N=3,\lambda_1=1)$, which holds for the SU(3) spin structure factor
\begin{equation}
{\cal O}^{jj}(\omega)  = \frac{1}{2\pi Z} \int_{-\infty}^{+\infty} dt 
                  e^{i\omega t} \Tr\left[e^{-\beta\hat{\mathcal{H}}} \hat{O}^{j}(t)\hat{O}^{j}(0) \right].
\label{eq:ojj_qm}
\end{equation}
A similar renormalization factor $s^{\alpha}_{\bm{r}} \to \sqrt{1+1/S} s^{\alpha}_{\bm{r}}$ must be applied to the classical spins obtained from SU(2) coherent states to fulfill the sum rule in the infinite temperature limit~\cite{Huberman_2008}. In general, the renormalization factor that must be applied to recover the sum rule for arbitrary values of $N$ and $\lambda_1$ is:
\begin{equation}
\kappa(N,\lambda_1) =\frac{ \sqrt{C_1 (N,\lambda_1)}}{ \sqrt{(N-1)^3/(2N)} \lambda_1}=
\sqrt{1+N/\lambda_1}
\end{equation}
where $C_1(N,\lambda_1)= (N-1)^3\lambda_1^2/(2N) + (N-1)^3\lambda_1/2 $ is the eigenvalue of the quadratic Casimir operator of SU($N$) for the degenerate irrep $[\lambda_1, 0, \ldots,0]$~\footnote{The eigenvalues of the Casimir operators for semi-simple Lie groups were derived by Perelomov and Popov in~\cite{PerelomovPopov_1968}. Note that the eigenvalue of the quadratic Casimir operator of SU($N$) in the physical basis considered in this work can be obtained by multiplying the eigenvalue in the basis adopted by Perelomov and Popov by a factor  $(N-1)^2/2$.}. 

In the low temperature limit, $\beta DS^2 \to \infty$, the dynamics is controlled by the normal modes of the quadratic fluctuations around the minimum energy classical state.  The intensity of each excited  mode is proportional to $T$ because of the equipartition theorem. The exact quantum mechanical result can be recovered by multiplying the classical dynamical spin structure factor by $\beta \omega$, which is a well-known quantum-classical correspondence  for the harmonic oscillator~\cite{LLD_SZhang,Huberman_2008}. Note also that the ground state of $\hat{\mathcal{H}}_{\rm SI}$ has no net dipole moment,
$\langle {\hat {\bm S}}\rangle=0$, for  $D>0$. In other words, the state has only a net quadrupolar moment, implying that the orbit of SU(2) coherent states is not enough to represent the classical limit of this state. This is another clear indicator of the need of using a bigger Lie algebra to define a classical limit of $\hat{\mathcal{H}}_{\rm SI}$ that captures the qualitative aspects of the exact quantum mechanical solution. 

In summary, the spin dynamics of $\hat{\mathcal{H}}_{\rm SI}$ is well approximated by a classical LLD based on 
SU(3) coherent states, but it is not well-described by the traditional SU(2) LLD. Assuming that $|D|$ is comparable or bigger than $z J$ ($z$ is the coordination number and $J$ is the characteristic energy scale of the exchange tensor), this statement holds true for  the full Hamiltonian $\hat{\mathcal{H}}$ [see Eq.~\eqref{eq:model}] for two simple reasons. In the high-$T$ limit, the dynamics of $\hat{\mathcal{H}}$ is well approximated by the dynamics of $\hat{\mathcal{H}}_{\rm SI}$. The basic role of the interaction term of $\hat{\mathcal{H}}$ is to broaden the delta function shown in Fig.~\eqref{fig:fig2}. In the low-$T$ limit, the ground state of $\hat{\mathcal{H}}$ is a quantum paramagnet with no net dipolar moment for $D>0$. As we already explained for the single-ion case, such a state has no classical counterpart within the orbit of SU(2) coherent states. We note that this statement remains true for the \emph{magnetically ordered state} if the system is relatively close the to quantum critical point that divides this phase from the quantum paramagnet~\cite{do2020decay}, implying that it is still necessary to use SU(3) coherent states to approximate the spin dynamics of the ordered magnet in the proximity of the QCP. Interestingly, this statement remains true for the easy-axis case $D<0$ and $|D|$ comparable or bigger than $z J$. In this limit, the ground state is an Ising-like magnetically ordered state, which definitely has classical counterpart in SU(2) coherent states. However, the low-energy modes of this classical ground state include quadrupolar fluctuations~\cite{bai2021hybridized}, which are not captured by the traditional SU(2) LLD.


\section{Conclusions}
\label{conc}

In this work we exploited the fact that an $N$-level quantum mechanical system admits more than one classical limit to generalize the LLD for spin systems. As it was noticed  by Perelomov~\citep{perelomov1972} and Gilmore~\citep{Gilmore1972391}, different classical limits can be obtained by introducing coherent states of different Lie algebras and generalizing Dirac's conjecture. A spin $S$ Hamiltonian admits more than one classical limit because it can be expressed as a function of generators of different Lie algebras~\cite{Batista04}. Traditionally, spin $S$ Hamiltonians are expressed as functions of the components of the physical spin, which are  generators of SU(2) in the spin $S$ irrep. However, one can express the same Hamiltonian as a function of generators of SU($2S+1$) in the fundamental representation~\cite{Batista04} and introduce a new classical limit based on coherent states of SU($2S+1$). While this is not the only alternative to the traditional classical limit of a spin $S$ system, it has multiple advantages that were discussed in the previous sections of this work.

A clear advantage is that the orbit of coherent states is maximized for SU$(N)$ if $N$ is the local dimension of the coherent state. Based on the variational principle, the ``SU($N$) classical limit'' provides  the best estimation of the ground state energy in comparison with other classical limits. For the same reason, SU($N$) coherent states allow us to describe physical states with no net dipolar moment (they are characterized by higher order multipoles) and coherent fluctuations that change the relative magnitude of the different multi-polar moments  (i.e. fluctuations that are different from physical rotations). A common example is provided by the longitudinal modes of ordered magnets that change the length of the dipole moment. For a spin one system, these modes correspond to SU(3) rotations that change the relative magnitudes of the dipolar and quadrupolar components~\cite{bai2021hybridized,do2020decay}. 

In general, the main advantage of introducing a classical limit to approximate the dynamics of interacting quantum spin systems is a huge reduction in computational cost. While the cost of computing the exact dynamics grows exponentially in the number of spins $N_s$, the cost of simulating the classical spin dynamics grows linearly in $N_s$. The only difference in computational cost between the two classical limits presented here is a prefactor $N^2-1$ associated with the number of generators of SU$(N)$ (there is one equation of motion for each generator). In view of the comparable computational cost and vast range of applications of the traditional LLD, we envision that the generalization proposed in this work will have an important impact on the modeling of large classes of magnets that either include a large  single-ion anisotropy (e.g. crystal field splitting of $d$ and $f$-electron magnets) or  consist of weakly-coupled multi-spin units (e.g. weakly coupled spin dimers, trimers or quadrumers).
In this work we used an example of the first case to illustrate the need of introducing the SU($N$) classical limit. To understand why weakly coupled multi-spin units require a similar treatment, it is enough to consider the simplest case of weakly-coupled $S=1/2$  dimers. The local Hilbert space has dimension four and the eigenstates of a single-dimer antiferromagnetic Heisenberg Hamiltonian are the singlet ground state and a triplet of degenerate excited states. Once again, SU($2$) coherent states are not enough to describe the non-magnetic character of the singlet state. In contrast, SU(4) coherent states allow us to describe a quantum paramagnet, as well as a magnetically ordered state (induced by a strong enough inter-dimer coupling) with a strong reduction of the magnitude of the ordered moment. Moreover, the SU(4) classical LLD accounts for the $N-1=3$ excited triplon modes of the quantum paramagnet   that become dispersive for finite inter-dimer coupling. Note that the classical limit that is obtained with SU(4) coherent states corresponds to the classical limit of the semi-classical expansion (or generalized spin wave theory)~\cite{Matsumoto04,Zapf14} that is constructed with the bond operators introduced by  Sachdev and Bhatt~\cite{Sachdev90}.

\begin{acknowledgments}
 We thank Xiaojian Bai, Kipton Barros, David Dahlbom, Ying Wai Li, Cole Miles, Shizeng Lin, Martin Mourigal, Matthew Scott Wilson, David Alan Tennant,  Zhentao Wang and Shang-Shun Zhang  for useful discussions.  H.~Z. was partially supported by the Shull Wollan Center Graduate Research Fellowship.  The work of C.D.B was supported by the U.S. Department of Energy, Office of Science, Basic Energy Sciences, Materials Sciences and Engineering Division under award No.~DE-SC-0018660.
\end{acknowledgments}

\appendix

\section{Expansion in SU(2) coherent states}
\label{apd:su2}
Consider the expectation value of the product of two on-site operators $\hat{A}$ and $\hat{B}$ 
for an SU(2) coherent state,
\begin{align}
AB(\theta,\phi) & = \langle \Omega(\theta,\phi)
                    |\hat{A}\hat{B}|\Omega(\theta,\phi)\rangle \nonumber\\
                & = \frac{2S+1}{4\pi}\int{\rm d}
                    \vec{\Omega}' |\langle\Omega|\Omega'\rangle|^{2}
                    \frac{\langle \Omega| \hat{A} |\Omega' \rangle}
                    {\langle\Omega|\Omega'\rangle}
                    \frac{\langle \Omega'|\hat{B}| \Omega\rangle}
                    {\langle\Omega'|\Omega\rangle},
\label{eq:absu2}
\end{align}
where we inserted the resolution of identity in terms of  SU(2) coherent
states, and
\begin{equation}
\frac{2S+1}{4\pi}d\vec{\Omega} = \frac{2S+1}{4\pi}d\phi d\theta\sin\theta
\end{equation}
is the Haar measure of SU(2). The first term of the above integrand is the overlap between two SU(2) coherent states~\citep{auerbach2012interacting}
\begin{equation}
|\langle\Omega|\Omega'\rangle|^{2}=\bigg(\frac{1+\vec{\Omega}\cdot\vec{\Omega}'}{2}\bigg)^{2S},
\end{equation}
where $\vec{\Omega}=(\sin\theta\cos\phi,\sin\theta\sin\phi,\cos\theta)$. In the large-$S$ limit [recall that $S=\lambda_1/2$ for SU(2)], the overlap becomes arbitrarily small except for $\vec{\Omega}'\simeq\vec{\Omega}$, i.e. two different SU(2) coherent states become orthogonal to each other. 
The large value of $S$ justifies a  \emph{saddle-point approximation} to evaluate the integral in Eq.~\eqref{eq:absu2}.
After expanding the term $\vec{\Omega}\cdot\vec{\Omega}'$ up to the quadratic order in $\delta\gamma=(\gamma'-\gamma)$, where $\gamma=\theta,\phi$,
\begin{align}
\vec{\Omega} \cdot \vec{\Omega}' & = \cos\theta\cos(\theta+\delta\theta)
                                   + \cos(\delta\phi)\sin(\theta)
                                     \sin(\theta+\delta\theta) \nonumber\\
                                 & = 1 - \frac{1}{2}(\delta\theta)^{2}
                                   - \frac{1}{2} \sin^{2}\theta (\delta\phi)^{2}
                                   + \mathcal{O}\big[(\delta\gamma)^{3}\big],
\end{align}
we rewrite the square of the overlap as
\begin{align}
|\langle\Omega|\Omega'\rangle|^{2} =e^{\ln|\langle\Omega|\Omega'\rangle|^{2}} 
 & \simeq e^{2S \ln\big[ 1-\frac{1}{4}(\delta\theta)^{2}
  -\frac{1}{4}\sin^{2}\theta(\delta\phi)^{2} \big]} \nonumber\\
 & \simeq e^{-S \big[(\theta'-\theta)^{2}/2 + \sin^{2}\theta(\phi'-\phi)^{2}/2 \big]}.
\end{align}
Since  (for fixed $\theta$ and $\phi$) the second factor and the third factor of the integrand in Eq.~\eqref{eq:absu2} are analytical functions of the complex variables $\alpha'=\sin\theta' \phi'+i\theta'$
and $\bar{\alpha}'=\sin\theta' \phi'-i\theta'$, respectively, we can expand the two terms as we did in Eqs.~\eqref{eq:papp} and \eqref{eq:ppbp},
\begin{widetext}
\begin{align}
AB(\theta,\phi) & \simeq \frac{2S+1}{4\pi}\int_{0}^{2\pi}d\phi'
                         \int_{0}^{\pi}d\theta' \sin\theta' \nonumber\\
                & \times e^{-S\big[ (\theta'-\theta)^{2}/2
                 + \sin^{2}\theta(\phi'-\phi)^{2}/2\big] }
                   \bigg\{ A(\theta,\phi)B(\theta,\phi) \nonumber\\
                & -\frac{i}{2} \bigg[ \frac{1}{\sin\theta}
                \frac{\partial A}{\partial\theta}
                \frac{\partial B}{\partial\phi}
                (\theta'-\theta)^{2}-\frac{1}{\sin\theta}
                \frac{\partial A}{\partial\phi}
                \frac{\partial B}{\partial\theta}
                (\sin\theta' \phi'-\sin\theta\phi)^{2} \bigg] \nonumber\\
                & +\frac{1}{2} \bigg[ \frac{1}{\sin\theta}
                \frac{\partial A}{\partial\theta}
                \frac{\partial B}{\partial\theta}(
                \theta'-\theta)^{2}
                +\frac{1}{\sin\theta}\frac{\partial A}{\partial\phi}
                \frac{\partial B}{\partial\phi}
                (\sin\theta' \phi'-\sin\theta\phi)^{2} \bigg]
                +\mathcal{L}\bigg\},
\end{align}
\end{widetext}
where $\mathcal{L}$ includes terms (up to quadratic order)  that vanish after the integration. Since the width of the Gaussian goes to zero in the large-$S$ limit, we can extend the integration limits to infinity and set $\sin\theta'=\sin\theta$ to recover the familiar form of
the Gaussian integration. In summary, we have
\begin{align}
AB(\theta,\phi) & = A(\theta,\phi)B(\theta,\phi)\big(1+1/(2S)\big)\nonumber \\
                & -\frac{i}{2}\frac{1}{S\sin\theta}
                \bigg[ \frac{\partial A}{\partial\theta}
                       \frac{\partial B}{\partial\phi}
                      -\frac{\partial A}{\partial\phi}\frac{\partial B}{\partial\theta}
                      \bigg] \nonumber \\
                & +\frac{1}{2}\frac{1}{S\sin\theta}
                \bigg[ \frac{\partial A}{\partial\theta}
                       \frac{\partial B}{\partial\theta}
                       +\frac{\partial A}{\partial\phi}
                       \frac{\partial B}{\partial\phi} \bigg]
                       +\mathcal{O}\big[(1/S)^{2}\big]
\end{align}
By taking the large-$S$ limit, we prove the factorization rule for
SU(2)
\begin{equation}
AB(\theta,\phi)\xrightarrow{S\rightarrow\infty} A(\theta,\phi)B(\theta,\phi).
\end{equation}
The above result also provides the definition of the Poisson bracket on the orbit of SU(2) coherent states
\begin{align}
\big\{ A(\theta,\phi),B(\theta,\phi)\}_{PB}
  & = -i\lim_{S\rightarrow\infty} S \big[ A,B \big](\theta,\phi)\nonumber \\
  & = \frac{1}{\sin\theta}\bigg( \frac{\partial A}{\partial\phi}
      \frac{\partial B}{\partial\theta}
     -\frac{\partial A}{\partial\theta}
      \frac{\partial B}{\partial\phi} \bigg).
\end{align}
Note that $\{\phi\sin\theta,\theta\}_{PB}=1$, implying that $\phi\sin\theta$ and $\theta$ play the role of canonical coordinate and momentum variables defined on the $S^2\simeq \text{CP}^{1}$ manifold of SU(2) coherent states. 
Finally, the SU(2) L-L equation Eq.~\eqref{eq:tradll} takes the following form in terms of these spherical coordinates
\begin{align}
\frac{{\rm d}\theta}{{\rm d}t} & = -\frac{1}{\sin\theta}
                                    \frac{\partial h}{\partial\phi} =\{\theta,h\}_{PB}\nonumber \\
\frac{{\rm d}\phi}{{\rm d}t}   & =  \frac{1}{\sin\theta}
                                    \frac{\partial h}{\partial\theta}=\{\phi,h\}_{PB},
\end{align}
where $h$ is the classical Hamiltonian.

\section{Physical basis of  generators of SU(3)\label{apd:su3}}

The ``physical'' basis of generators of SU(3) is obtained by applying the following transformation to  the natural basis  defined in Eq.~\eqref{eq:sun_basis}:
\begin{align}
\hat{O}_{1} & =\hat{S}^{x}=\frac{1}{\sqrt{2}}(\hat{g}_{13}+\hat{g}_{31}+\hat{g}_{23}+\hat{g}_{32}),\nonumber \\
\hat{O}_{2} & =\hat{S}^{y}=\frac{1}{\sqrt{2}}(-i\hat{g}_{13}+i\hat{g}_{31}+i\hat{g}_{23}-i\hat{g}_{32}),\nonumber \\
\hat{O}_{3} & =\hat{S}^{z}=2\hat{H}_{1},\nonumber \\
\hat{O}_{4} & =-(\hat{S}^{x}\hat{S}^{z}+\hat{S}^{z}\hat{S}^{x})=\frac{1}{\sqrt{2}}(-\hat{g}_{13}-\hat{g}_{31}+\hat{g}_{23}+\hat{g}_{32}),\nonumber \\
\hat{O}_{5} & =-(\hat{S}^{y}\hat{S}^{z}+\hat{S}^{z}\hat{S}^{y})=\frac{1}{\sqrt{2}}(i\hat{g}_{13}-i\hat{g}_{31}+i\hat{g}_{23}-i\hat{g}_{32}),\nonumber \\
\hat{O}_{6} & =(\hat{S}^{x})^{2}-(\hat{S}^{y})^{2}=(\hat{g}_{12}+\hat{g}_{21}),\nonumber \\
\hat{O}_{7} & =\hat{S}^{x}\hat{S}^{y}+\hat{S}^{y}\hat{S}^{x}=-i(\hat{g}_{12}-\hat{g}_{21}),\nonumber \\
\hat{O}_{8} & =\sqrt{3}(\hat{S}^{z})^{2}-\frac{2}{\sqrt{3}}=\frac{2}{\sqrt{3}}(\hat{H}_{1}+2\hat{H}_{2}),
\label{eq:su3_physicalop}
\end{align}
where $\hat{O}_{1-3}$ are the three components of the vector known as dipole moment  and $\hat{O}_{4-8}$ are the five components of the quadrupolar moment, which is a symmetric traceless  tensor of rank two. The commutation relations between the generators of SU(3) can be formally written as 
\begin{equation}
\big[\hat{O}_{a},\hat{O}_{b}\big]=i \sum_{c}f_{abc}\hat{O}_{c},\ a,b,c=1,\ldots,8,
\label{eq:su3_commu}
\end{equation}
where the structure coefficients $f_{abc}$ are completely anti-symmetric in the three indices, generalizing the Levi-Civita symbol of SU(2). The non-zero coefficients (with zero permutation) take the values
\begin{align}
 & f_{123}=1,\ f_{147}=1,\ f_{156}=-1,\ f_{158}=-\sqrt{3},\nonumber \\
 & f_{246}=-1,\ f_{248}=\sqrt{3},\ f_{257}=-1,\ f_{345}=1,\ f_{367}=2.
 \label{eq:su3_strconst}
\end{align}

\section{Correlation function in the classical limit\label{apd:corrcl}}

In this appendix we consider the transverse dynamical spin structure factor of the single-ion problem Eq.~\eqref{eq:sion}  to illustrate how to compute dynamical correlation functions in the classical limit based on SU($N$) coherent states. The trace in Eq.~\eqref{eq:spm_qm} now runs over the CP$^{N-1}$ orbit of SU($N$) coherent states
\begin{multline}
    \mathcal{S}^{+-}(\omega) = \frac{1}{2\pi} \int_{-\infty}^{+\infty} d t e^{i\omega t} 
    \frac{1}{Z} \int \mathcal{D}[\{\alpha_\rho\}]  \\
    \times \langle \Omega[\{\alpha_\rho\}] | e^{-\beta\hat{\mathcal{H}}}\hat{S}^{+}(t) \hat{S}^-(0) | \Omega [\{\alpha_\rho\}] \rangle,
\end{multline}
where $\mathcal{D}[\{\alpha_\rho\}]$ is the integration measure that can be found in Ref.~\cite{Nemoto_2000} and the partition function is 
\begin{equation}
    Z = \int \mathcal{D}[\{\alpha_\rho\}] \langle \Omega[\{\alpha_\rho\}] | e^{-\beta\hat{\mathcal{H}}} | \Omega [\{\alpha_\rho\}] \rangle.
\end{equation}
By  exploiting the factorization rule given in Eq.~\eqref{eq:factorization}, which is valid in the classical limit, we obtain
\begin{align}
   \mathcal{S}^{+-}(\omega) & \simeq \frac{1}{2\pi} \int_{-\infty}^{+\infty} d t e^{i\omega t} 
    \frac{1}{Z} \int \mathcal{D}[\{\alpha_\rho\}] e^{-\beta h[\{\alpha_\rho\}]}  \nonumber \\
    & \times \lim_{\tau \rightarrow\infty} \frac{1}{T} \int_{-\tau/2}^{+\tau/2}d t' s^+(t+t') (\{\alpha_\rho\}) s^-(t') (\{\alpha_\rho\}) \nonumber \\
    & = \lim_{\tau \rightarrow\infty}\frac{\tau}{2\pi Z} \int \mathcal{D}[\{\alpha_\rho\}] e^{-\beta h[\{\alpha_\rho\}]} s^+(\omega)s^-(-\omega)\label{eq:clsqw},
\end{align}
where $h[\{\alpha_\rho\}]=\langle \Omega[\{\alpha_\rho\}]|\hat{H}|| \Omega [\{\alpha_\rho\}] \rangle$ is the classical Hamiltonian and  we used the convolution theorem along with the time translation symmetry of ${\hat {\cal H}}$. The quantity
\begin{equation}
    s^+(\omega) =  \lim_{\tau \rightarrow \infty} \frac{1}{\tau} \int_{-\infty}^{+\infty}d t e^{i\omega t} s^+(t)
\end{equation}
is the Fourier transform of the classical spin operator $s^+(t)$ and  $s^-(-\omega)=[s^+(\omega)]^*$. Note that  Eq.~\eqref{eq:clsqw} provides the basis for numerical implementations of the traditional LLD (e.g. Refs.~\citep{LLD_SZhang,LLD_AS_honeycomb,LLD_AS_kitaev}): the trace over the SU($N$) coherent states is usually computed by applying the Metropolis-Hastings Monte Carlo algorithm and $T$ is a finite time much longer than any characteristic time of the system.


\bibliographystyle{apsrev4-2}
%
\end{document}